\documentstyle[12pt]{article}
\begin{document}
\noindent
\begin{center}
{\Large {\bf Gravitational Coupling and\\ Dynamical Reduction of
The Cosmological Constant \\}} \vspace{2cm}

 ${\bf Yousef~
Bisabr}$\footnote{e-mail:~y-bisabr@srttu.edu.}\\
\vspace{.5cm} {\small{Department of Physics, Shahid Rajaee
University,
Lavizan, Tehran 16788, Iran.}}\\
\end{center}
\vspace{1cm}
\begin{abstract}
We introduce a dynamical model to reduce a large cosmological
constant to a sufficiently small value. The basic ingredient in
this model is a distinction which has been made between the two
unit systems used in cosmology and particle physics.  We have used
a conformal invariant gravitational model to define a particular
conformal frame in terms of large scale properties of the
universe.  It is then argued that the contributions of mass scales
in particle physics to the vacuum energy density should be
considered in a different conformal frame.  In this manner, a
decaying mechanism is presented in which the conformal factor
appears as a dynamical field and plays a key role to relax a large
effective cosmological constant. Moreover, we argue that this
model also provides a possible explanation for the coincidence
problem.

\end{abstract}
\vspace{1cm} ~~~~~~~~PACS: 98.80.-k, 95.36.+x, 98.80.Es.
 \vspace{3cm}
\section{Introduction}
There are now strong observational evidences that the expansion of
the universe is accelerating. These observations are based on type
Ia supernova \cite{super} and Cosmic Microwave Background
Radiation \cite{cmbr}.  The standard explanation invokes an
unknown component, usually referred to as dark energy.  It
contributes to energy density of the universe with
$\Omega_{d}=0.7$ where $\Omega_{d}$ is the corresponding density
parameter, see e.g., \cite{ca} and references therein.  A
candidate for dark energy which seems to be both natural and
consistent with observations is the cosmological constant
\cite{ca} \cite{wein} \cite{cc}. However, in order to avoid
theoretical problems \cite{wein}, other scenarios have been
investigated. Among these scenarios, there are quintessence
\cite{q}, tachyons \cite{tach}, phantom \cite{ph},
quintom \cite{qui}, modified gravity \cite{modi} and so forth.\\
There are two important problems that are related to the
cosmological constant.  The first problem, usually known as the
fine tuning problem, is the large discrepancy between observations
and theoretical predictions on its value \cite{wein}. The second
problem concerns with the coincidence between the observed vacuum
energy density and the current matter density.  While these two
energy components evolve differently as the universe expands,
their contributions to total energy density of the universe in the
present epoch are the same order of magnitude.  In the present
work our main focus will be on the first problem and some comments
will be offered concerning the second
one.\\
There have been many attempts trying to resolve the fine tuning
problem \cite{wein}. Most of them are based on the belief that the
cosmological constant may not have such an extremely small value
at all times and there should exist a dynamical mechanism working
during evolution of the universe which provides a cancelation of
the vacuum energy density at late times \cite{cancel}.  As noted
in \cite{bis}, such a mechanism should have two important
characteristics.  Firstly, since any mass scale in particle
physics contributes to vacuum energy density much larger than the
observational bound, the mechanism should not be sensitive to a
particular type of contribution and should work equally well for
every mass scale introduced by elementary particle physics.
Secondly, it should work whenever these contributions are
considered at cosmological level since the discrepancy manifests
when one compares them with relevant cosmological observations.
This latter strongly suggests that construction of a mechanism for
relaxing these contributions should somehow take into account the
two unit systems usually used in cosmology and particle physics.
In fact, the small upper limit is obtained in a unit system which
is defined in terms of large scale cosmological parameters.  On
the other hand, the theoretical predictions are based on a natural
unit system which is suggested by quantum physics. These two unit
systems are usually related by a conversion factor which is
independent of space and time.  In other terms they are related by
a global unit transformation.  The point we wish to make here is
that the large discrepancy between observations and theoretical
predictions on the vacuum energy density arises when one
prejudices that the two unit systems should be indistinguishable
up to a constant conversion factor in all spacetime points, or
related by a global unit transformation. Such a global
transformation clearly carries no dynamical implications and the
use of a particular unit system is actually a matter of
convenience.  It means that one may arbitrarily use the unit
system suggested by quantum physics to describe the evolution of
the universe or the cosmological unit system to describe
dynamical properties of an elementary particle.\\
In the present work we would like to consider a theoretical scheme
in which an explicit recognition is given to the distinguished
characteristics of these two unit systems.  In such a theoretical
scheme one should no longer accept the triviality one usually
assigns to a unit transformation.  In this respect, we shall
consider local unit transformations (conformal transformations
\cite{3}) relating different standard of units (conformal frames)
via general spacetime dependent conversion (conformal) factors. In
this language observations and theoretical predictions on the
vacuum energy density are actually carried out in two different
conformal frames.  We emphasize that local unit transformations
give a dynamical meaning to changes of unit systems and can be
consequently taken as a basis for constructing a dynamical
mechanism which works due to cosmic expansion.  Along this line of
investigation, we shall show that the conformal factor which
relates the two unit systems plays the role of a dynamical field
which can eventually reduce the effective cosmological constant to
a small value consistent with observations. We also show that this
dynamical reduction of the vacuum energy density can also lead to
a possible alleviation of the
coincidence problem.  \\
Throughout this paper we work in units in which $\hbar=c=1$ and
the sign conventions are those of MTW \cite{5}.

\section{The Model}
We begin with considering the general form of vacuum sector of a
scalar tensor theory
\begin{equation}
S=\frac{1}{2} \int d^{4}x \sqrt{-g}~
\{F(\phi)~R+Z(\phi)~g^{\mu\nu} \nabla_{\mu}\phi
\nabla_{\nu}\phi+V(\phi)\}\label{a}\end{equation} Here $g$ is the
determinant of $g_{\mu\nu}$ and $R$ is the scalar curvature. The
functions F($\phi$), Z($\phi$) and V($\phi$) are arbitrary
functions of the real scalar field $\phi$.  Different
parameterizations have been used in the literature. However, we
would like to consider the case that the action (\ref{a}) remains
invariant under conformal transformations
\begin{equation}
\bar{g}_{\mu\nu} =e^{-2\sigma}  g_{\mu\nu} \label{c}\end{equation}
\begin{equation}
\bar{\phi} = e^{\sigma}  \phi \label{d}\end{equation}
 where
$\sigma$ is a smooth dimensionless spacetime function.  The
conformal transformation (\ref{c}) implies that all spacetime
intervals transform according to
\begin{equation}
\bar{ds}=e^{-\sigma} ds \label{ds}\end{equation} while coordinates
are fixed. This can be interpreted as changes of standards of
length and time or transformation of unit systems \cite{3}. In
this view under a conformal transformation all dimensional
quantities are transformed according to their dimensions.
Therefore if one assigns a mass $\mu$ to the scalar field $\phi$,
it should be transformed as
\begin{equation}
\bar{\mu}=e^{\sigma}\mu \label{e}\end{equation} With this fact in
mind, the parametrization $F(\phi)=\frac{1}{6}\phi^2$, $Z(\phi)=1$
and $V(\phi)=-\frac{1}{3}\mu^{2}\phi^{2}$ reduces the action
(\ref{a}) to \footnote{One may consider a self-interacting scalar
field by adding the term $\lambda \phi^{4}$ to the potential
$V(\phi)$. This would not change the conformal symmetry of the
action (\ref{b}).}
\begin{equation}
S=\frac{1}{2} \int d^{4}x \sqrt{-g}~ \{\frac{1}{6} R
\phi^{2}+g^{\mu\nu} \nabla_{\mu}\phi
\nabla_{\nu}\phi-\frac{1}{3}\mu^{2}\phi^{2}\}
\label{b}\end{equation} This action with $\mu=0$ is the field
theoretic version of the so-called Hoyle-Narlikar theory
\cite{nar}.  It is also shown \cite{pal} that $f(R)$ theories of
gravity \cite{modi} in the Palatini formalism can be cast in the
form of (\ref{b}) with an
appropriate potential.\\
The action (\ref{b}) is invariant under transformations (\ref{c}),
(\ref{d}) and (\ref{e}). Since Noether's theorem relates a basic
symmetry to a conservation law, two different conservation laws
can be attributed to the gravitational system (\ref{b}). Firstly,
general covariance requires that (\ref{b}) remains unchanged under
general coordinate transformations while the system of units is
held fixed.   This is related to the Bianchi identities or
conservation of the stress tensor of any matter system which
couples with (\ref{b}). Secondly, conformal invariance leads to a
relation between the trace of the stress tensor and $\mu$
\cite{deser}.  In the case that $\mu=o$, only
traceless matter systems can be coupled with (\ref{b}).\\
The conformal invariance of (\ref{b}) is broken if one assigns a
particular constant value to $\mu$. In this way one characterizes
the unit system in which a particular measurement is carried out.
In a cosmological context, the most suggestive choice for $\mu$ is
$\mu \sim H_{0}$ with $H_{0}^{-1}$ being the present Hubble radius
of the universe. In the corresponding conformal frame, which is
referred from now on as the cosmological frame, a constant
configuration can be assigned to the scalar field. It is given by
$\phi^{-2} \sim G$ with $G$ being the gravitational constant
\cite{deser}. The action (\ref{b}) then reduces to\footnote{For
arguments concerning the physical status of Einstein and Jordan
frames see, for example, \cite{capo}.}
\begin{equation}
S=\frac{1}{\kappa} \int d^{4}x \sqrt{-g} (R-2\mu^2)
\label{f}\end{equation} where $\kappa$ gives the gravitational
constant. This corresponds to the usual Einstein-Hilbert action
with a small but nonzero cosmological constant induced due to
finite size of the universe. However, the cosmological constant
receives strong contributions from various mass scales introduced
by elementary particle physics. To incorporate these contributions
to (\ref{f}) it should be noted that they belong to a conformal
frame which has properties entirely different from those used to
define the cosmological frame. To define this atomic conformal
frame or unit system, one considers local characteristics of a
typical elementary particle and neglects the large scale
properties of the universe \cite{bis} \footnote{In fact, the basic
idea that a viable cosmological model should contain a dynamical
distinction between the unit systems used in cosmology and
particle physics is not new. See, for example, \cite{canuto}.}. If
the metric tensor of this conformal frame is denoted by
$\bar{g}_{\mu\nu}$, it is then related to $g_{\mu\nu}$ by
(\ref{c}). We therefore write the action (\ref{f}) in the form
\begin{equation}
S=\frac{1}{\kappa} \int d^{4}x \sqrt{-g} (R-2\mu^2) -\int d^{4}x
\sqrt{-\bar{g}} L(\bar{g}_{\mu\nu},
\bar{\psi})\label{h}\end{equation} where $L(\bar{g}_{\mu\nu},
\bar{\psi})$ is the Lagrangian density of some matter field
$\bar{\psi}$ in the atomic conformal frame. As an illustration we
take $L(\bar{g}_{\mu\nu}, \bar{\psi})$ to be Lagrangian density of
a real massive scalar field
\begin{equation}
L(\bar{g}_{\mu\nu},
\bar{\psi})=\bar{g}^{\mu\nu}\nabla_{\mu}\bar{\psi}\nabla_{\nu}\bar{\psi}+\bar{m}^2\bar{\psi}^2
\label{g}\end{equation} The quantity $\bar{m}$ corresponds to the
mass of the scalar field in the atomic unit system.  In terms of
the background variables ($g_{\mu\nu}$, $\psi$), (\ref{g}) can be
written as
\begin{equation}
L(\bar{g}_{\mu\nu},
\bar{\psi})=e^{4\sigma}\{g^{\mu\nu}\nabla_{\mu}\psi\nabla_{\nu}\psi+
\psi^{2}g^{\mu\nu}\nabla_{\mu}\sigma\nabla_{\nu}\sigma +2\psi
 g^{\mu\nu}\nabla_{\mu}\psi\nabla_{\nu}\sigma+\bar{m}^2e^{-2\sigma}\psi^2\}
\label{i}\end{equation} If we suppose that the scalar field $\psi$
has a constant average value in the cosmological frame we can set
$\psi$=constant.  In this case (\ref{i}) reduces to
\begin{equation}
L(\bar{g}_{\mu\nu},
\bar{\psi})=\psi^2e^{4\sigma}\{g^{\mu\nu}\nabla_{\mu}\sigma\nabla_{\nu}\sigma
+\bar{m}^2e^{-2\sigma}\} \label{j}\end{equation} Combining
(\ref{h}) and (\ref{j}) leads to
\begin{equation}
S= \int d^{4}x \sqrt{-g} \{R-2\mu^2-\kappa\alpha(
g^{\mu\nu}\nabla_{\mu}\sigma\nabla_{\nu}\sigma+\bar{m}^2
e^{-2\sigma}) \}\label{k}\end{equation} where $\psi^2=\alpha$.
Note that $\sigma$ appears as a dynamical field which allows us to
investigate the evolution of $\bar{m}$ as the universe evolves. We
now follow the consequences of the action (\ref{k}) by writing the
field equations
\begin{equation}
G_{\mu\nu}+\mu^2 g_{\mu\nu}=\kappa~ T^{\sigma}_{\mu\nu}
\label{l}\end{equation}
\begin{equation}
\Box \sigma+\bar{m}^2 e^{-2\sigma}=0 \label{m}\end{equation} where
\begin{equation}
T^{\sigma}_{\mu\nu}=\alpha (\nabla_{\mu}\sigma
\nabla_{\nu}\sigma-\frac{1}{2}g_{\mu\nu} \nabla^{\gamma}\sigma
\nabla_{\gamma}\sigma-\frac{1}{2}\bar{m}^2 e^{-2\sigma}
g_{\mu\nu}) \label{ll}\end{equation}
 In
these equations the exponential coefficient for $\bar{m}^2$
emphasizes the dynamical distinction between the two unit systems
mentioned above.  In an expanding universe this distinction is
expected to increase since cosmological scales enlarge as the
universe expands and, as suggested by (\ref{ds}), the conformal
factor $e^{\sigma}$ must grow with time.  This authomatically
provides us with a dynamical reduction of $\bar{m}$ in the
cosmological frame. That this intuitive picture is actually
consistent with the field equations is illustrated in
the following:\\
Applying (\ref{l}) and (\ref{m}) to the spatially flat
Friedmann-Robertson-Walker metric, yields
\begin{equation}
3 H^2-\mu^2=\kappa \rho_{\sigma} \label{n}\end{equation}
\begin{equation}
2\dot{H}+3H^2 -\mu^2=-\kappa p_{\sigma}\label{n1}\end{equation}
\begin{equation}
\ddot{\sigma}+3H \dot{\sigma}-\bar{m}^2e^{-2\sigma}=0
\label{o}\end{equation} where
\begin{equation}
\rho_{\sigma}=\frac{1}{2} \alpha (\dot{\sigma}^2 +\bar{m}^2
e^{-2\sigma}) \label{o1}\end{equation}
\begin{equation}
p_{\sigma}=\frac{1}{2}\alpha (\dot{\sigma}^2 -\bar{m}^2
e^{-2\sigma}) \label{o2}\end{equation} Here $H=\frac{\dot{a}}{a}$
is the Hubble parameter and overdot indicates differentiation with
respect to the coordinate time $t$. Due to homogeneity and
isotropy, the field $\sigma$ is taken to be only a function of
time. Assuming that the universe follows a power law expansion,
namely that $H \sim t^{-1}$, equation (\ref{o}) gives the solution
\begin{equation}
e^{\sigma}=\sigma_{0} t \label{oo}\end{equation}  with $\sigma_{0}
\sim \bar{m}$. In equation (\ref{n}), one then obtains
$\Lambda_{eff} \equiv \mu^2 -\bar{m}^2 e^{-2\sigma} \sim t^{-2}$
which is consistent with the observational bound.\\
The other problem attributed to the cosmological constant is the
coincidence between matter and vacuum energy densities in total
energy density of the universe.  In standard $\Lambda$CDM model
energy density of matter dilutes as $\rho_{m} \sim a^{-3}$ while
vacuum energy density is a constant. Thus one should explain that
in the history of the universe why we live in an epoch in which
these two energy densities are of the same order of magnitude. One
possible explanation is that the vacuum energy density is not
actually a constant and evolves during expansion of the universe.
If the evolution of the vacuum density is the same as that of the
matter density we then arrive at a possible solution of the
problem.\footnote{One may note that the model presented here
shares some likeness with the quintessence models.  However,
quintessence models need appropriate potential functions for a
scalar field which are usually constructed by fine tuning.
Moreover, there is no a precise physical meaning which can be
attributed to the scalar field.} Let us now examine this property
in the model presented here. To do this we should add a cosmic
matter system to the action (\ref {k}),
\begin{equation}
S= \int d^{4}x \sqrt{-g} \{(R-2\mu^2) +\kappa[ L_{m}(g_{\mu\nu})-
\alpha( g^{\mu\nu}\nabla_{\mu}\sigma\nabla_{\nu}\sigma+\bar{m}^2
e^{-2\sigma})]\}\label{hh}\end{equation} where $L_{m}$ denotes
lagrangian density corresponding to the cosmic matter. Note that
this cosmological matter couples with the metric $g_{\mu\nu}$
describing the cosmological frame.  The equation (\ref{l}) then
generalizes to
\begin{equation}
G_{\mu\nu}+\mu^2 g_{\mu\nu}=\kappa~
(T^{m}_{\mu\nu}+T^{\sigma}_{\mu\nu}) \label{lll}\end{equation}
where $T^{m}_{\mu\nu}$ is the stress tensor corresponding to the
cosmological matter system.  Applying the Bianchi identities and
using the relation (\ref{ll}), one can easily check that the
stress tensors $T^{\sigma}_{\mu\nu}$ and $T^{m}_{\mu\nu}$ are
separately conserved
\begin{equation}
\nabla^{\mu}T^{\sigma}_{\mu\nu}= \nabla^{\mu}T^{m}_{\mu\nu}=0
\end{equation}
We take the cosmic matter to be a dust system (perfect fluid with
zero pressure) with energy density $\rho_{m}$. In this case the
equation (\ref{n}) takes the form
\begin{equation}
3 H^2-\mu^2=\kappa(\rho_{m} +\rho_{\sigma})
\label{nn}\end{equation} while (\ref{n1}) and (\ref{o}) remain
unchanged. We then define the ratio
$r\equiv\frac{\rho_{m}}{\rho_{\sigma}}$ and use the equations
(\ref{n1}) and (\ref{nn}) to obtain
\begin{equation} r=-1-\frac{2\dot{H}}{\kappa
\rho_{\sigma}}-\frac{p_{\sigma}}{\rho_{\sigma}}
\label{r}\end{equation} This relation implies that $r$ remains
constant during expansion of the universe if $p_{\sigma}$ and
$\rho_{\sigma}$ evolves as $\dot{H}$.  If the universe follows a
power law expansion then $H\sim t^{-1}$.  In this case, matter and
vacuum densities have the same time evolution if $p_{\sigma}\sim
\rho_{\sigma}\sim t^{-2}$.  Inspection of (\ref{o1}) and
(\ref{o2}) reveals that this condition is  actually consistent
with the decaying law of $\bar{m}$ which is represented by
(\ref{oo}).\\
As the last point we remark that different couplings of matter
systems with gravity, as indicated in (\ref{hh}), may seem to be
in conflict with equivalence principle or universality of free
fall. However, it should be noted that this principle is supported
by very precise experiments which have been carried out in the
present epoch and it is not clear that it does hold during
evolution of the universe. Noting the relation (\ref{oo}), one
infers that the dynamical field $\sigma$ which makes the
gravitational coupling of cosmic matter and atomic matter to be
different during evolution of the universe takes at present time a
constant value $\bar{m} e^{-\sigma}\sim H_{0}^{-1}$. This implies
the same coupling of the two types of matter systems in the
present epoch if $\alpha \sim 1$.
\section{Conclusion}
We argued that the large discrepancy between observations and
theoretical estimations on vacuum energy density may be attributed
to an interrelation between the unit systems by which these
quantities are usually measured. We have proposed  a dynamical
model in which these two unit systems are related by a local unit
transformation.  The basic ingredient in our model is the
gravitational coupling of different contributions to vacuum energy
density coming from elementary particle physics. These
contributions belong to a conformal frame that is dynamically
distinct from the cosmological frame in which the cosmological
observations are carried out. Mathematically, it means that these
contributions couple with a metric $\bar{g}_{\mu\nu}$ which is
conformally related to the metric defined in the cosmological
frame $g_{\mu\nu}$ by a spacetime dependent conformal factor. We
would like to underline two important features of such a
gravitational coupling. Firstly, different mass scales introduced
by particle physics have variable contributions to vacuum energy
density in the cosmological frame. secondly, these variations are
controlled by the conformal factor. Note that this conformal
factor is characterized by $\sigma$ field which automatically
finds a kinetic term in the action (\ref{k}) when we write a
particular field theory, denoted here by $L(\bar{g}_{\mu\nu},
\bar{\psi})$, in terms of the background variables. From a
physical point of view, one expects that $\sigma$ be an increasing
function of time since due to expansion of the universe the
cosmological scales should enlarge with respect to the atomic
scales. This then provides us with a dynamical reduction of the
cosmological constant which works due to cosmic expansion. As the
last point, we remark that decaying of the cosmological constant
with expansion of the universe provides a possible explanation for
the coincidence problem. Alleviating the problem needs the same
time evolution for both matter and vacuum energy densities.  We
have shown that this behavior is actually consistent with our
field equations.


\end{document}